\documentclass[a4paper,aps,onecolumn,nofootinbib]{revtex4}
\RequirePackage[colorlinks,hyperindex]{hyperref}
\RequirePackage[english]{babel}
\RequirePackage[latin1]{inputenc}
\RequirePackage[T1]{fontenc}
\RequirePackage{mathrsfs}
\RequirePackage{amsmath}
\RequirePackage{amssymb}
\RequirePackage{amsbsy}
\RequirePackage{color}
\RequirePackage{bm}
\hypersetup{colorlinks=true,breaklinks=true,urlcolor=blue,linkcolor=red}
\begin{document}
\title{\bf{Classical characters of spinor fields in torsion gravity}}
\author{Luca Fabbri$^{\nabla}$\!\!\! $^{\hbar}$\footnote{luca.fabbri@unige.it}}
\affiliation{$^{\nabla}$DIME, Universit\`{a} di Genova, Via all'Opera Pia 15, 16145 Genova, ITALY\\
$^{\hbar}$INFN, Sezione di Genova, Via Dodecaneso 33, 16146 Genova, ITALY}
\date{\today}
\begin{abstract}
We consider the problem of having relativistic quantum mechanics re-formulated with hydrodynamic variables, and specifically the problem of deriving the Mathisson-Papapetrou-Dixon equations from the Dirac equation. The problem will be answered on a general manifold with torsion and gravity. We will demonstrate that when plane waves are considered the MPD equations acquire the form given in \cite{Guedes:2022mdy}, but we will also see that in such a form the MPD equations become trivial.
\end{abstract}
\maketitle
\section{Introduction}
More than a century ago, the passage from classical to quantum physics marked the beginning of an era of unprecedented scientific and technological achievements that still continues today. The price we must pay for this revolution is the abandonment of the visualizability of physical concepts. In the passage from classical to quantum mechanics, our capacity to master the laws of nature has increased, but we have lost the ability to see what we are doing (this is the state of affairs that pushed famous figures of the past to say that no one really understands quantum mechanics).

In such a situation, two philosophical currents arose: one suggesting to disregard any interest about the meaning so long as the correct numbers were computed (as exemplified by D. Mermin's notorious 'shut up and calculate'); and one believing that our comprehension is improved if we can attribute an interpretation to those numbers. The adherents to the second philosophical current are those who advocate a return to the visualizability of classical concepts.

Of all attempts to have quantum mechanics re-expressed with classical pictures, the de Broglie-Bohm interpretation\footnote{Recall that in dBB mechanics, hidden variables are non-local to be compatible with the violation of Bell-like inequalities [J. F. Clauser, M. A. Horne, A. Shimony, R. A. Holt, ``Proposed experiment to test local hidden variable theories'', \textit{Phys. Rev. Lett.} \textbf{23}, 880 (1969)].} is the one that best implements the program by having quantum mechanics re-formulated in terms of hydrodynamic variables: the idea is that the wave function be re-written as the product of a module times a phase, which are then identified with density and velocity. The Schr\"{o}dinger equation correspondingly decomposes into one Hamilton-Jacobi equation and one continuity equation, and the full hydrodynamic formulation is then accomplished. The extension to include spin and to enforce relativistic invariance has been source of debate between Bohm himself and Takabayasi in references \cite{b1, t1, b2, t2}. Mathematical details have been addressed in \cite{jl1, jl2}. See also \cite{Book} for a more recent overview.

Along this direction, the next step is to ask if, from the fundamental equations of relativistic quantum mechanics, we can derive the Newton equation. Or better, because relativistic quantum fields have spin, we should ask if we can obtain the Mathisson-Papapetrou-Dixon equations \cite{M, P, D}. As the MPD equations are obtained from the conservation laws of energy and spin, which naturally couple to curvature and torsion \cite{S, K}, it also becomes relevant to ask what happens when space-times possessing non-trivial structures are considered \cite{HKH}. A first answer has been given in \cite{OK} for the Weyssenhoff fluid. A more realistic answer was worked out in \cite{Guedes:2022mdy} for Dirac fields in plane waves.

Here, we will give the general answer to this problem for the case of Dirac fields without restraining ourselves to a plane wave solution. We will show however that when plane waves are considered the MPD equations as given in \cite{Guedes:2022mdy} are recovered, although we will also see that in such a form the MPD equations reduce to be trivial.
\section{Dirac Field in Polar Form}\label{general}
\subsection{Spinor Kinematics}
\subsubsection{Algebra}
Let $\boldsymbol{\gamma}^{i}$ be matrices in the Clifford algebra, from which $\boldsymbol{\sigma}_{ik}\!=\![\boldsymbol{\gamma}_{i},\boldsymbol{\gamma}_{k}]/4$ are the generators of the complex Lorentz group. From $2i\boldsymbol{\sigma}_{ab}\!=\!\varepsilon_{abcd}\boldsymbol{\pi}\boldsymbol{\sigma}^{cd}$ we implicitly define the parity-odd matrix $\boldsymbol{\pi}$ telling that the complex Lorentz group is reducible (this matrix is usually denoted by a gamma with an index five, which we will suppress since it is not a true index). Exponentiation of the generators gives the complex Lorentz group $\boldsymbol{\Lambda}$ and therefore $\boldsymbol{S}\!=\!\boldsymbol{\Lambda}e^{iq\alpha}$ gives the spinor group. Spinor fields are objects transforming as $\psi\!\rightarrow\!\boldsymbol{S}\psi$ and $\overline{\psi}\!\rightarrow\!\overline{\psi}\boldsymbol{S}^{-1}$ where $\overline{\psi}\!=\!\psi^{\dagger}\boldsymbol{\gamma}^{0}$ is the adjoint operation: with a pair of adjoint spinors one forms the spinorial bi-linear quantities
\begin{eqnarray}
&\Sigma^{ab}\!=\!2\overline{\psi}\boldsymbol{\sigma}^{ab}\boldsymbol{\pi}\psi\ \ \ \
\ \ \ \ \ \ \ \ M^{ab}\!=\!2i\overline{\psi}\boldsymbol{\sigma}^{ab}\psi\label{tensors}\\
&S^{a}\!=\!\overline{\psi}\boldsymbol{\gamma}^{a}\boldsymbol{\pi}\psi\ \ \ \
\ \ \ \ \ \ \ \ U^{a}\!=\!\overline{\psi}\boldsymbol{\gamma}^{a}\psi\label{vectors}\\
&\Theta\!=\!i\overline{\psi}\boldsymbol{\pi}\psi\ \ \ \
\ \ \ \ \ \ \ \ \Phi\!=\!\overline{\psi}\psi\label{scalars},
\end{eqnarray}
all of which being real tensors, with parity-odd and parity-even counter-parts in the left and right columns. We have
\begin{eqnarray}
&M_{ab}\Phi\!-\!\Sigma_{ab}\Theta\!=\!U^{j}S^{k}\varepsilon_{jkab}\ \ \ \ \ \ \ \
\ \ \ \ \ \ \ \ M_{ab}\Theta\!+\!\Sigma_{ab}\Phi\!=\!U_{[a}S_{b]}\label{MSigma}
\label{momenta}
\end{eqnarray}
as well as
\begin{eqnarray}
&M_{ik}U^{i}=\Theta S_{k}\ \ \ \ \ \ \ \ \Sigma_{ik}U^{i}\!=\!\Phi S_{k}\label{products-1}\\
&M_{ik}S^{i}=\Theta U_{k}\ \ \ \ \ \ \ \ \Sigma_{ik}S^{i}\!=\!\Phi U_{k}\label{products-2}
\end{eqnarray}
together with the orthogonality relations
\begin{eqnarray}
&\frac{1}{2}M_{ab}M^{ab}\!=\!-\frac{1}{2}\Sigma_{ab}\Sigma^{ab}\!=\!\Phi^{2}\!-\!\Theta^{2} \label{orthogonal}\\
&\frac{1}{2}M_{ab}\Sigma^{ab}\!=\!-2\Theta\Phi\label{norm}
\end{eqnarray}
and
\begin{eqnarray}
&U_{a}U^{a}\!=\!-S_{a}S^{a}\!=\!\Theta^{2}\!+\!\Phi^{2}\label{NORM}\\
&U_{a}S^{a}\!=\!0\label{ORTHOGONAL},
\end{eqnarray}
known Fierz re-arrangements. We also have $\Sigma^{ab}\!=\!-\frac{1}{2}\varepsilon^{abij}M_{ij}$ and by combining both conditions in (\ref{MSigma}) we get
\begin{eqnarray}
&M_{ab}(\Theta^{2}\!+\!\Phi^{2})\!=\!\Phi U^{j}S^{k}\varepsilon_{jkab}\!+\!\Theta U_{[a}S_{b]}
\label{M}
\end{eqnarray}
showing that if $\Theta^{2}\!+\!\Phi^{2}\!\neq\!0$ all the bi-linears are writable in terms of the two vectors and the two scalars. Under this condition (which is generally verified), spinor fields can always be written, in chiral representation, in the polar form
\begin{eqnarray}
&\psi\!=\!\phi\ e^{-\frac{i}{2}\beta\boldsymbol{\pi}}
\ \boldsymbol{L}^{-1}\left(\begin{tabular}{c}
$1$\\
$0$\\
$1$\\
$0$
\end{tabular}\right)
\label{spinor}
\end{eqnarray}
for a pair of functions $\phi$ and $\beta$ and for some $\boldsymbol{L}$ with the structure of a spinor transformation \cite{jl1, jl2, Book}. With it, we get
\begin{eqnarray}
&\Theta\!=\!2\phi^{2}\sin{\beta}\ \ \ \ \ \ \ \ \ \ \ \ \ \ \ \ \Phi\!=\!2\phi^{2}\cos{\beta}
\end{eqnarray}
showing that $\beta$ and $\phi$ are a pseudo-scalar and a scalar, called chiral angle and density. Then we can introduce
\begin{eqnarray}
&S^{a}\!=\!2\phi^{2}s^{a}\ \ \ \ \ \ \ \ \ \ \ \ \ \ \ \ U^{a}\!=\!2\phi^{2}u^{a}
\end{eqnarray}
as the spin axial-vector and velocity vector, in terms of which (\ref{NORM}-\ref{ORTHOGONAL}) reduce to
\begin{eqnarray}
&u_{a}u^{a}\!=\!-s_{a}s^{a}\!=\!1\\
&u_{a}s^{a}\!=\!0
\end{eqnarray}
showing that the velocity has $3$ independent components, the $3$ spatial rapidities, whereas the spin has only $2$ independent components, the $2$ angles that, in the rest-frame, its spatial part forms with one given axis. Therefore, $\boldsymbol{L}$ is the specific transformation that takes the assigned spinor field to its rest-frame spin-eigenstate.
\subsubsection{Differentiation}
Differential structures for spinor fields are defined upon the introduction of a gauge potential and a spin connection having torsion \cite{S, K}. In the present treatment we will consider torsion to be completely antisymmetric since this is the only irreducible component of torsion that can be excited by spin-$1/2$ particles such as the Dirac field (for more detailed discussions see \cite{Fabbri:2017lmf}). Since any completely antisymmetric tensor as a Hodge dual, we can write
\begin{eqnarray}
&W^{\mu}\!=\!Q_{\alpha\sigma\nu}\varepsilon^{\alpha\sigma\nu\mu}
\label{W}
\end{eqnarray}
where $W^{\mu}$ is called torsion axial-vector. The non-symmetric affine connection can be decomposed as
\begin{eqnarray}
&\Gamma^{\alpha}_{\mu\nu}\!=\!\Lambda^{\alpha}_{\mu\nu}
\!+\!\frac{1}{2}Q^{\alpha}_{\phantom{\alpha}\mu\nu}
\label{Q}
\end{eqnarray}
where $\Lambda^{\alpha}_{\mu\nu}$ is the unique symmetric connection that can always be written in terms of the partial derivatives of the metric, known as Levi-Civita connection (we will always work under the condition of metric-compatibility).

Our aim next is to apply the general covariant derivation of spinors to the case in which spinors are in polar form: the torsionfull covariant derivative of spinor fields is given by
\begin{eqnarray}
&\boldsymbol{D}_{\mu}\psi\!=\!\partial_{\mu}\psi\!+\!\boldsymbol{\Omega}_{\mu}\psi
\label{covder}
\end{eqnarray}
where
\begin{eqnarray}
&\boldsymbol{\Omega}_{\mu}\!=\!
\frac{1}{2}\Omega_{ab\mu}\boldsymbol{\sigma}^{ab}\!+\!iqA_{\mu}\mathbb{I}
\label{spinor-conn}
\end{eqnarray}
in which $A_{\mu}$ is the gauge potential of charge $q$ and $\Omega_{ab\mu}$ is the torsionfull spin connection \cite{Fabbri:2017lmf}. For polar spinors
\begin{eqnarray}
&\boldsymbol{D}_{\mu}\psi
\!=\!(D_{\mu}\ln{\phi}\mathbb{I}\!-\!\frac{i}{2}D_{\mu}\beta\boldsymbol{\pi}
\!+\!\partial_{\mu}\boldsymbol{L}^{-1}\boldsymbol{L}\!+\!iqA_{\mu}\mathbb{I}
\!+\!\frac{1}{2}\Omega_{ij\mu}\boldsymbol{\sigma}^{ij})\psi
\end{eqnarray}
after easy computations. As the logarithmic derivative of elements of a Lie group belongs to its Lie algebra, we have
\begin{eqnarray}
\boldsymbol{L}^{-1}\partial_{\mu}\boldsymbol{L}\!=\!iq\partial_{\mu}\xi\mathbb{I}
\!+\!\frac{1}{2}\partial_{\mu}\xi^{ab}\boldsymbol{\sigma}_{ab}
\label{gold}
\end{eqnarray}
for some $\partial_{\mu}\xi$ and $\partial_{\mu}\xi^{ab}$ known as Goldstone fields \cite{Fabbri:2021mfc}. Then we can write
\begin{eqnarray}
&\boldsymbol{D}_{\mu}\psi
\!=\!(D_{\mu}\ln{\phi}\mathbb{I}\!-\!\frac{i}{2}D_{\mu}\beta\boldsymbol{\pi}
\!-\!iP_{\mu}\mathbb{I}
\!-\!\frac{1}{2}G_{ij\mu}\boldsymbol{\sigma}^{ij})\psi
\label{covder-polar}
\end{eqnarray}
where the quantities
\begin{eqnarray}
&\partial_{\mu}\xi_{ij}\!-\!\Omega_{ij\mu}\!:=\!G_{ij\mu}\label{G}\\
&q(\partial_{\mu}\xi\!-\!A_{\mu})\!:=\!P_{\mu}\label{P}
\end{eqnarray}
were introduced. To see that $G_{ij\mu}$ and $P_{\mu}$ are real tensors, consider a generic spinor transformation $\boldsymbol{S}$, acting on the spinor field $\psi\!\rightarrow\!\boldsymbol{S}\psi$: because its covariant derivative (\ref{covder}) also transforms as $\boldsymbol{D}_{\mu}\psi\!\rightarrow\!\boldsymbol{S}\boldsymbol{D}_{\mu}\psi$ then
\begin{eqnarray}
\boldsymbol{\Omega}_{\mu}\!\rightarrow\!\boldsymbol{S}\left(\boldsymbol{\Omega}_{\mu}
-\boldsymbol{S}^{-1}\partial_{\mu}\boldsymbol{S}\right)\boldsymbol{S}^{-1}\label{connection};
\end{eqnarray}
simultaneously, when $\psi\!\rightarrow\!\boldsymbol{S}\psi$ is applied to a spinor in polar form we have that
\begin{eqnarray}
&\phi\ e^{-\frac{i}{2}\beta\boldsymbol{\pi}}
\ \boldsymbol{L}^{-1}\left(\begin{tabular}{c}
$1$\\
$0$\\
$1$\\
$0$
\end{tabular}\right)\!\rightarrow\!\boldsymbol{S}\phi\ e^{-\frac{i}{2}\beta\boldsymbol{\pi}}
\ \boldsymbol{L}^{-1}\left(\begin{tabular}{c}
$1$\\
$0$\\
$1$\\
$0$
\end{tabular}\right)
\end{eqnarray}
and because $\phi$ and $\beta$ are scalars, $\phi\!\rightarrow\!\phi$ and $\beta\!\rightarrow\!\beta$, and then we obtain
\begin{eqnarray}
\boldsymbol{L}^{-1}\!\rightarrow\!\boldsymbol{S}\boldsymbol{L}^{-1}\label{L}.
\end{eqnarray}
Putting (\ref{connection}) and (\ref{L}) together gives
\begin{eqnarray}
&\partial_{\mu}\boldsymbol{L}^{-1}\boldsymbol{L}\!+\!\boldsymbol{\Omega}_{\mu}
\!\rightarrow\!\partial_{\mu}(\boldsymbol{L}\boldsymbol{S}^{-1})^{-1}(\boldsymbol{L}\boldsymbol{S}^{-1})\!+\!\boldsymbol{S}\!\left(\boldsymbol{\Omega}_{\mu}
\!-\!\boldsymbol{S}^{-1}\partial_{\mu}\boldsymbol{S}\right)\!\boldsymbol{S}^{-1}
\!=\!\boldsymbol{S}\left(\partial_{\mu}\boldsymbol{L}^{-1}\boldsymbol{L}
\!+\!\boldsymbol{\Omega}_{\mu}\right)\boldsymbol{S}^{-1}
\end{eqnarray}
showing that $\partial_{\mu}\boldsymbol{L}^{-1}\boldsymbol{L}\!+\!\boldsymbol{\Omega}_{\mu}$ transforms as a spinorial matrix. Decomposing it with (\ref{spinor-conn}) and (\ref{gold}) yields
\begin{eqnarray}
\nonumber
&\partial_{\mu}\boldsymbol{L}^{-1}\boldsymbol{L}\!+\!\boldsymbol{\Omega}_{\mu}
\!=\!-\left(\frac{1}{2}\partial_{\mu}\xi^{ab}\boldsymbol{\sigma}_{ab}
\!+\!iq\partial_{\mu}\xi\mathbb{I}\right)
\!+\!\left(\frac{1}{2}\Omega_{ab\mu}\boldsymbol{\sigma}^{ab}\!+\!iqA_{\mu}\mathbb{I}\right)=\\
&=-\frac{1}{2}(\partial_{\mu}\xi_{ab}\!-\!\Omega_{ab\mu})\boldsymbol{\sigma}^{ab}
\!-\!iq(\partial_{\mu}\xi\!-\!A_{\mu})\mathbb{I}
\!=\!-\frac{1}{2}G_{ab\mu}\boldsymbol{\sigma}^{ab}\!-\!iP_{\mu}\mathbb{I}:
\label{spinorialmatrixdec}
\end{eqnarray}
the linear independence of $\boldsymbol{\sigma}^{ab}$ and $\mathbb{I}$ lets us split
\begin{eqnarray}
&(G_{ab\mu}\boldsymbol{\sigma}^{ab})'
\!=\!G_{ab\mu}\boldsymbol{\Lambda}\boldsymbol{\sigma}^{ab}\boldsymbol{\Lambda}^{-1}\\
&P_{\mu}'\!=\!P_{\mu}
\end{eqnarray}
and as the complex representation $\boldsymbol{\Lambda}$ and the real representation $(\Lambda)^{a}_{i}$ are linked by $(\boldsymbol{\sigma}^{ab})'\!=\!\boldsymbol{\Lambda}\boldsymbol{\sigma}^{ij}\boldsymbol{\Lambda}^{-1}(\Lambda)^{a}_{i}(\Lambda)^{b}_{j}$ then
\begin{eqnarray}
&G'_{ab\mu}\!=\!G_{ij\mu}(\Lambda^{-1})^{i}_{a}(\Lambda^{-1})^{j}_{b}
\end{eqnarray}
showing that $G_{ab\mu}$ and $P_{\mu}$ are a real tensor and a gauge-covariant vector, called space-time tensorial connection and gauge tensorial connection \cite{Fabbri:2021mfc}. From (\ref{covder-polar}), we can compute
\begin{eqnarray}
&\overline{\psi}\boldsymbol{\gamma}^{k}\boldsymbol{D}_{\mu}\psi
\!=\!U^{k}D_{\mu}\ln{\phi}\!-\!\frac{i}{2}D_{\mu}\beta S^{k}\!-\!iP_{\mu}U^{k}
\!+\!\frac{i}{4}G_{ij\mu}\varepsilon^{kijq}S_{q}-\frac{1}{2}G_{ij\mu}\eta^{ki}U^{j}
\end{eqnarray}
and thus, adding its complex conjugate, leads to
\begin{eqnarray}
&\boldsymbol{D}_{\mu}(\overline{\psi}\boldsymbol{\gamma}_{k}\psi)
\!=\!2U_{k}D_{\mu}\ln{\phi}+U^{j}G_{jk\mu}
\end{eqnarray}
in terms of the velocity alone. This can be written as
\begin{eqnarray}
&2D_{\mu}\phi^{2}u_{k}+2\phi^{2}D_{\mu}u_{k}
\!=\!2\phi^{2}u_{k}D_{\mu}\ln{\phi^{2}}+2\phi^{2}u^{j}G_{jk\mu}
\end{eqnarray}
where the derivatives of the density cancel. Eventually $D_{\mu}u_{k}\!=\!u^{j}G_{jk\mu}$. The same could be proven for the spin. So
\begin{eqnarray}
&D_{\mu}s_{k}\!=\!G_{jk\mu}s^{j}\ \ \ \ \ \ \ \
\ \ \ \ \ \ \ \ D_{\mu}u_{k}\!=\!G_{jk\mu}u^{j}
\end{eqnarray}
showing that the tensor $G_{ab\mu}$ is related to the covariant derivatives of velocity and spin.

At the second-order derivative, the commutator is given by
\begin{eqnarray}
&[\boldsymbol{D}_{\mu},\boldsymbol{D}_{\nu}]\psi
\!=\!Q^{\alpha}_{\phantom{\alpha}\mu\nu}\boldsymbol{D}_{\alpha}\psi
\!+\!\boldsymbol{G}_{\mu\nu}\psi
\label{spinorcommutator}
\end{eqnarray}
in which
\begin{eqnarray}
&\boldsymbol{G}_{\mu\nu}\!=\!\frac{1}{2}G_{ab\mu\nu}\boldsymbol{\sigma}^{ab}
\!+\!iqF_{\mu\nu}\mathbb{I}\label{curvstrengdec}
\end{eqnarray}
is the decomposition of the curvature in terms of the torsionfull Riemann curvature $G_{ab\mu\nu}$ and the Maxwell strength $F_{\mu\nu}$ \cite{Fabbri:2017lmf}. These can be written in terms of space-time and gauge tensorial connections (\ref{G}-\ref{P}): in fact, (\ref{covder-polar}) implies
\begin{eqnarray}
\nonumber
&\boldsymbol{D}_{\mu}\boldsymbol{D}_{\nu}\psi
\!=\!D_{\mu}(D_{\nu}\ln{\phi}\mathbb{I}\!-\!\frac{i}{2}D_{\nu}\beta\boldsymbol{\pi}
\!-\!iP_{\nu}\mathbb{I}
\!-\!\frac{1}{2}G_{ij\nu}\boldsymbol{\sigma}^{ij})\psi+\\
\nonumber
&+(D_{\nu}\ln{\phi}\mathbb{I}\!-\!\frac{i}{2}D_{\nu}\beta\boldsymbol{\pi}
\!-\!iP_{\nu}\mathbb{I}
\!-\!\frac{1}{2}G_{ij\nu}\boldsymbol{\sigma}^{ij})(D_{\mu}\ln{\phi}\mathbb{I}\!-\!\frac{i}{2}D_{\mu}\beta\boldsymbol{\pi}
\!-\!iP_{\mu}\mathbb{I}
\!-\!\frac{1}{2}G_{ij\mu}\boldsymbol{\sigma}^{ij})\psi=\\
\nonumber
&=(D_{\mu}D_{\nu}\ln{\phi}\mathbb{I}\!-\!\frac{i}{2}D_{\mu}D_{\nu}\beta\boldsymbol{\pi}
\!-\!iD_{\mu}P_{\nu}\mathbb{I}
\!-\!\frac{1}{2}D_{\mu}G_{ij\nu}\boldsymbol{\sigma}^{ij}+\\
\nonumber
&+D_{\nu}\ln{\phi}D_{\mu}\ln{\phi}\mathbb{I}
\!-\!\frac{i}{2}D_{\mu}\ln{\phi}D_{\nu}\beta\boldsymbol{\pi}
\!-\!iP_{\nu}D_{\mu}\ln{\phi}\mathbb{I}
\!-\!\frac{1}{2}D_{\mu}\ln{\phi}G_{ij\nu}\boldsymbol{\sigma}^{ij}-\\
\nonumber
&-\frac{i}{2}D_{\nu}\ln{\phi}D_{\mu}\beta\boldsymbol{\pi}
\!-\!\frac{1}{4}D_{\nu}\beta D_{\mu}\beta\mathbb{I}
\!-\!\frac{1}{2}P_{\nu}D_{\mu}\beta\boldsymbol{\pi}
\!+\!\frac{i}{4}D_{\mu}\beta G_{ij\nu}\boldsymbol{\sigma}^{ij}\boldsymbol{\pi}-\\
\nonumber
&-iP_{\mu}D_{\nu}\ln{\phi}\mathbb{I}
\!-\!\frac{1}{2}P_{\mu}D_{\nu}\beta\boldsymbol{\pi}
\!-\!P_{\nu}P_{\mu}\mathbb{I}
\!+\!\frac{i}{2}P_{\mu}G_{ij\nu}\boldsymbol{\sigma}^{ij}-\\
&-\frac{1}{2}D_{\nu}\ln{\phi}G_{ij\mu}\boldsymbol{\sigma}^{ij}
\!+\!\frac{i}{4}D_{\nu}\beta G_{ij\mu}\boldsymbol{\sigma}^{ij}\boldsymbol{\pi}
\!+\!\frac{i}{2}P_{\nu}G_{ij\mu}\boldsymbol{\sigma}^{ij}
\!+\!\frac{1}{4}G_{cd\nu}G_{ab\mu}\boldsymbol{\sigma}^{cd}\boldsymbol{\sigma}^{ab})\psi
\end{eqnarray}
so that its antisymmetrization yields
\begin{eqnarray}
\nonumber
&[\boldsymbol{D}_{\mu},\boldsymbol{D}_{\nu}]\psi
\!=\!-[-Q^{\alpha}_{\phantom{\alpha}\mu\nu}D_{\alpha}\ln{\phi}\mathbb{I}
\!+\!\frac{i}{2}Q^{\alpha}_{\phantom{\alpha}\mu\nu}D_{\alpha}\beta\boldsymbol{\pi}
\!+\!i(D_{\mu}P_{\nu}\!-\!D_{\nu}P_{\mu})\mathbb{I}+\\
&+\frac{1}{2}(D_{\mu}G_{ij\nu}\!-\!D_{\nu}G_{ij\mu}
\!+\!G_{ik\mu}G^{k}_{\phantom{k}j\nu}
\!-\!G_{ik\nu}G^{k}_{\phantom{k}j\mu})\boldsymbol{\sigma}^{ij}]\psi
\end{eqnarray}
in which $[\boldsymbol{\sigma}_{ab},\boldsymbol{\sigma}_{cd}]
=\eta_{ad}\boldsymbol{\sigma}_{bc}\!-\!\eta_{ac}\boldsymbol{\sigma}_{bd}
\!+\!\eta_{bc}\boldsymbol{\sigma}_{ad}\!-\!\eta_{bd}\boldsymbol{\sigma}_{ac}$ was used. From (\ref{spinorcommutator}) and (\ref{curvstrengdec}), and again (\ref{covder-polar}), we have
\begin{eqnarray}
\nonumber
&-[-Q^{\alpha}_{\phantom{\alpha}\mu\nu}D_{\alpha}\ln{\phi}\mathbb{I}
\!+\!\frac{i}{2}Q^{\alpha}_{\phantom{\alpha}\mu\nu}D_{\alpha}\beta\boldsymbol{\pi}
\!+\!i(D_{\mu}P_{\nu}\!-\!D_{\nu}P_{\mu})\mathbb{I}
\!+\!\frac{1}{2}(D_{\mu}G_{ij\nu}\!-\!D_{\nu}G_{ij\mu}
\!+\!G_{ik\mu}G^{k}_{\phantom{k}j\nu}
\!-\!G_{ik\nu}G^{k}_{\phantom{k}j\mu})\boldsymbol{\sigma}^{ij}]\psi=\\
&=[\boldsymbol{D}_{\mu},\boldsymbol{D}_{\nu}]\psi\!=\!(Q^{\alpha}_{\phantom{\alpha}\mu\nu}D_{\alpha}\ln{\phi}\mathbb{I}
\!-\!Q^{\alpha}_{\phantom{\alpha}\mu\nu}\frac{i}{2}D_{\alpha}\beta\boldsymbol{\pi}
\!-\!Q^{\alpha}_{\phantom{\alpha}\mu\nu}iP_{\alpha}\mathbb{I}
\!-\!Q^{\alpha}_{\phantom{\alpha}\mu\nu}\frac{1}{2}G_{ij\alpha}\boldsymbol{\sigma}^{ij}
\!+\!\frac{1}{2}G_{ab\mu\nu}\boldsymbol{\sigma}^{ab}\!+\!iqF_{\mu\nu}\mathbb{I})\psi
\end{eqnarray}
or equivalently
\begin{eqnarray}
\nonumber
&[i(D_{\mu}P_{\nu}\!-\!D_{\nu}P_{\mu})\mathbb{I}
\!-\!Q^{\alpha}_{\phantom{\alpha}\mu\nu}iP_{\alpha}\mathbb{I}
\!+\!iqF_{\mu\nu}\mathbb{I}+\\
&+\frac{1}{2}(D_{\mu}G_{ij\nu}\!-\!D_{\nu}G_{ij\mu}
\!+\!G_{ik\mu}G^{k}_{\phantom{k}j\nu}
\!-\!G_{ik\nu}G^{k}_{\phantom{k}j\mu})\boldsymbol{\sigma}^{ij}
\!-\!Q^{\alpha}_{\phantom{\alpha}\mu\nu}\frac{1}{2}G_{ij\alpha}\boldsymbol{\sigma}^{ij}
\!+\!\frac{1}{2}G_{ab\mu\nu}\boldsymbol{\sigma}^{ab}]\psi\!=\!0
\end{eqnarray}
after some re-arrangement. All the terms can now be grouped together in two major terms, one proportional to $\mathbb{I}$ and one proportional to $\boldsymbol{\sigma}^{ab}$, which are linearly independent. Therefore they have to be independently equal to zero, as
\begin{eqnarray}
&G^{i}_{\phantom{i}j\mu\nu}\!=\!-(D_{\mu}G^{i}_{\phantom{i}j\nu}
\!-\!D_{\nu}G^{i}_{\phantom{i}j\mu}
\!+\!G^{i}_{\phantom{i}k\mu}G^{k}_{\phantom{k}j\nu}
\!-\!G^{i}_{\phantom{i}k\nu}G^{k}_{\phantom{k}j\mu})
\!+\!G^{i}_{\phantom{i}j\alpha}Q^{\alpha}_{\phantom{\alpha}\mu\nu}\\
&qF_{\mu\nu}\!=\!-(D_{\mu}P_{\nu}\!-\!D_{\nu}P_{\mu})
\!+\!P_{\alpha}Q^{\alpha}_{\phantom{\alpha}\mu\nu}
\end{eqnarray}
which are the desired expressions of the torsionfull Riemann curvature and the Maxwell strength written in terms of space-time and gauge tensorial connections. Notice the torsional spurious term at the end of both.
\subsection{Spinor Dynamics}
The possibility to have the torsionfull connection $\Gamma^{\alpha}_{\mu\nu}$ decomposed into the torsionless connection $\Lambda^{\alpha}_{\mu\nu}$ plus torsion as in (\ref{Q}) means that this decomposition can be done on all quantities defined with the connection. So, for example, the covariant derivative applied to the case of a vector becomes
\begin{eqnarray}
&D_{\mu}V^{\alpha}\!=\!\partial_{\mu}V^{\alpha}\!+\!V^{\nu}\Gamma^{\alpha}_{\nu\mu}
\!=\!\partial_{\mu}V^{\alpha}\!+\!V^{\nu}(\Lambda^{\alpha}_{\nu\mu}
\!+\!\frac{1}{2}Q^{\alpha}_{\phantom{\alpha}\nu\mu})
\!=\!(\partial_{\mu}V^{\alpha}\!+\!V^{\nu}\Lambda^{\alpha}_{\nu\mu})
\!+\!\frac{1}{2}V^{\nu}Q^{\alpha}_{\phantom{\alpha}\nu\mu}:
\end{eqnarray}
since $\Lambda^{\alpha}_{\mu\nu}$ is a connection then
\begin{eqnarray}
&\nabla_{\mu}V^{\alpha}\!=\!\partial_{\mu}V^{\alpha}\!+\!V^{\nu}\Lambda^{\alpha}_{\nu\mu}
\end{eqnarray}
is also a covariant derivative. Using it in the previous expression gives
\begin{eqnarray}
&D_{\mu}V^{\alpha}\!=\!\nabla_{\mu}V^{\alpha}
\!+\!\frac{1}{2}V^{\nu}Q^{\alpha}_{\phantom{\alpha}\nu\mu}\label{decV}.
\end{eqnarray}
Analogously, the torsionfull covariant derivative of spinors $\boldsymbol{D}_{\mu}\psi$ can be decomposed in terms of the corresponding torsionless covariant derivative of spinors $\boldsymbol{\nabla}_{\mu}\psi$ plus torsional contributions as
\begin{eqnarray}
&\boldsymbol{D}_{\mu}\psi\!=\!\boldsymbol{\nabla}_{\mu}\psi
\!+\!\frac{1}{4}Q_{ab\mu}\boldsymbol{\sigma}^{ab}\psi\label{decSpin}.
\end{eqnarray}
In an identical manner, the torsionfull form of the space-time tensorial connection $G_{ab\mu}$ can be decomposed into the torsionless space-time tensorial connection $R_{ab\mu}$ plus torsion as
\begin{eqnarray}
&G_{ij\mu}\!=\!R_{ij\mu}\!-\!\frac{1}{2}Q_{ij\mu}\label{R}
\end{eqnarray}
and the torsionfull Riemann curvature $G_{ab\mu\nu}$ can be decomposed into the torsionless Riemann curvature $R_{ab\mu\nu}$ plus torsional terms as
\begin{eqnarray}
&G^{\sigma}_{\phantom{\sigma}\kappa\alpha\beta}
\!=\!R^{\sigma}_{\phantom{\sigma}\kappa\alpha\beta}
\!+\!\frac{1}{2}(\nabla_{\alpha}Q^{\sigma}_{\phantom{\sigma}\kappa \beta}
\!-\!\nabla_{\beta}Q^{\sigma}_{\phantom{\sigma}\kappa \alpha})
\!+\!\frac{1}{4}(Q^{\sigma}_{\phantom{\sigma}\rho \alpha}Q^{\rho}_{\phantom{\rho}\kappa \beta}
\!-\!Q^{\sigma}_{\phantom{\sigma}\rho \beta}Q^{\rho}_{\phantom{\rho}\kappa \alpha})\label{decG}.
\end{eqnarray}
The relevant torsionless quantities, when written using the polar variables, are given by
\begin{eqnarray}
&\boldsymbol{\nabla}_{\mu}\psi\!=\!(\nabla_{\mu}\ln{\phi}\mathbb{I}
\!-\!\frac{i}{2}\nabla_{\mu}\beta\boldsymbol{\pi}
\!-\!\frac{1}{2}R_{ij\mu}\boldsymbol{\sigma}^{ij}
\!-\!iP_{\mu}\mathbb{I})\psi
\label{decspinder}
\end{eqnarray}
and
\begin{eqnarray}
&\nabla_{\mu}s_{\nu}\!=\!s^{\alpha}R_{\alpha\nu\mu}\ \ \ \ \ \ \ \
\ \ \ \ \ \ \ \ \nabla_{\mu}u_{\nu}\!=\!u^{\alpha}R_{\alpha\nu\mu}\label{ds-du}
\end{eqnarray}
for the covariant derivatives of the spinor field and its bi-linears with
\begin{eqnarray}
&-R^{i}_{\phantom{i}j\mu\nu}\!=\!\nabla_{\mu}R^{i}_{\phantom{i}j\nu}
\!-\!\nabla_{\nu}R^{i}_{\phantom{i}j\mu}\!+\!R^{i}_{\phantom{i}k\mu}R^{k}_{\phantom{k}j\nu}
\!-\!R^{i}_{\phantom{i}k\nu}R^{k}_{\phantom{k}j\mu}\\
&-qF_{\mu\nu}\!=\!\nabla_{\mu}P_{\nu}\!-\!\nabla_{\nu}P_{\mu}\label{Maxwell}
\end{eqnarray}
for the torsionless Riemann curvature and Maxwell strength. Notice that no spurious term is present.

We now have all elements to discuss the dynamics, which is assigned by providing the Lagrangian.

From the torsionfull Riemann curvature one can define the contraction $G^{\alpha}_{\phantom{\alpha}\rho\alpha\sigma}\!=\!G_{\rho\sigma}$ called torsionfull Ricci tensor, and then $G^{\rho}_{\phantom{\rho}\rho}\!=\!G$ called torsionfull Ricci scalar. Analogously, $R^{\alpha}_{\phantom{\alpha}\rho\alpha\sigma}\!=\!R_{\rho\sigma}$ is the torsionless Ricci tensor, and $R^{\rho}_{\phantom{\rho}\rho}\!=\!R$ is the torsionless Ricci scalar. These are the scalars with which the Lagrangian can be built.

The simplest Lagrangian for the space-time dynamics is $\mathscr{L}\!=\!-G$ or explicitly
\begin{eqnarray}
&\mathscr{L}\!=\!-G\!\equiv\!-R\!+\!\frac{1}{4}Q^{\sigma\rho\beta}Q_{\sigma\rho\beta}
\end{eqnarray}
and its variation with respect to metric and torsion gives the Einstein equations, coupling the curvature to the energy density, and the Sciama-Kibble equations, coupling torsion to the spin density of the mater distribution \cite{S, K}. But nonetheless, since torsion is a tensor, there is no need to limit oneself to the torsional contribution that is implicitly defined in terms of the Ricci scalar, and additional square-torsion terms can be added. These supplementary terms will sum to the one already present, effectively shifting the constant in front of it. The new Lagrangian is thus
\begin{eqnarray}
&\mathscr{L}\!=\!-R\!+\!KQ^{\sigma\rho\beta}Q_{\sigma\rho\beta}
\!\equiv\!-R\!-\!\frac{1}{6}KW^{\mu}W_{\mu}
\end{eqnarray}
where $K$ is a generic constant. A further generalization can be obtained when considering that, as any physical field, torsion should propagate, and hence, derivative torsion terms should be allowed. This leads us to
\begin{eqnarray}
&\mathscr{L}\!=\!-R\!-\!\frac{1}{4}(\partial W)_{\alpha\nu}(\partial W)^{\alpha\nu}
\!+\!\frac{1}{2}M^{2}W^{\mu}W_{\mu}
\end{eqnarray}
where $(\partial W)_{\alpha\nu}\!=\!\nabla_{\alpha}W_{\nu}\!-\!\nabla_{\nu}W_{\alpha}$ and in which $M$ is the mass of the torsion axial-vector. The derivative term is given by the curl of the torsion axial-vector to ensure that the torsion dynamics be that of a Proca field and all the relative signs are fixed to ensure that mass and energy be real and positive. The inclusion of electrodynamics and spinor fields is done according to
\begin{eqnarray}
&\mathscr{L}\!=\!-R\!-\!\frac{1}{4}(\partial W)^{2}\!+\!\frac{1}{2}M^{2}W^{2}\!-\!\frac{1}{4}F^{2}
\!+\!\frac{i}{2}(\overline{\psi}\boldsymbol{\gamma}^{\mu}\boldsymbol{\nabla}_{\mu}\psi
\!-\!\boldsymbol{\nabla}_{\mu}\overline{\psi}\boldsymbol{\gamma}^{\mu}\psi)\!-\!m\Phi
\!-\!XW_{\nu}S^{\nu}
\end{eqnarray}
where $X$ is the torsion-spin coupling constant. This Lagrangian is the most general compatible with the requirement that the torsion axial-vector be a massive Proca field sourced by the Dirac spin axial-vector \cite{Fabbri:2017lmf}.

Its variation furnishes the torsion field equations
\begin{eqnarray}
&\nabla_{\rho}(\partial W)^{\rho\mu}\!+\!M^{2}W^{\mu}\!=\!XS^{\mu}
\label{torsion}
\end{eqnarray}
which are indeed the Proca equations (specifically, their divergence contains only lower-order derivatives). As for the gravitational field equations, they are
\begin{eqnarray}
\nonumber
&R^{\rho\sigma}\!-\!\frac{1}{2}Rg^{\rho\sigma}
\!=\!\frac{1}{2}[\frac{1}{4}F^{2}g^{\rho\sigma}
\!-\!F^{\rho\alpha}\!F^{\sigma}_{\phantom{\sigma}\alpha}+\\
\nonumber
&+\frac{1}{4}(\partial W)^{2}g^{\rho\sigma}
\!-\!(\partial W)^{\sigma\alpha}(\partial W)^{\rho}_{\phantom{\rho}\alpha}
\!+\!M^{2}(W^{\rho}W^{\sigma}\!-\!\frac{1}{2}W^{2}g^{\rho\sigma})+\\
&+\frac{i}{4}(\overline{\psi}\boldsymbol{\gamma}^{\rho}\boldsymbol{\nabla}^{\sigma}\psi
\!-\!\boldsymbol{\nabla}^{\sigma}\overline{\psi}\boldsymbol{\gamma}^{\rho}\psi
\!+\!\overline{\psi}\boldsymbol{\gamma}^{\sigma}\boldsymbol{\nabla}^{\rho}\psi
\!-\!\boldsymbol{\nabla}^{\rho}\overline{\psi}\boldsymbol{\gamma}^{\sigma}\psi)
\!-\!\frac{1}{2}X(W^{\sigma}S^{\rho}\!+\!W^{\rho}S^{\sigma})]
\label{gravitation}
\end{eqnarray}
in which the right-hand side is the energy (in particular, the torsion and electrodynamic contributions have a positive time-time component). The electrodynamic field equations are
\begin{eqnarray}
&\nabla_{\sigma}F^{\sigma\mu}\!=\!qU^{\mu}
\label{electrodynamics}
\end{eqnarray}
and the spinor field equation is
\begin{eqnarray}
&i\boldsymbol{\gamma}^{\mu}\boldsymbol{\nabla}_{\mu}\psi
\!-\!XW_{\sigma}\boldsymbol{\gamma}^{\sigma}\boldsymbol{\pi}\psi\!-\!m\psi\!=\!0.
\label{matter}
\end{eqnarray}
The Dirac field equation can be written with polar variables by substituting in (\ref{matter}) the expression (\ref{decspinder}): so
\begin{eqnarray}
\left[(\nabla_{\mu}\beta\!-\!2XW_{\mu}\!+\!B_{\mu})\boldsymbol{\gamma}^{\mu}\boldsymbol{\pi}
\!+\!i(\nabla_{\mu}\ln{\phi^{2}}\!+\!R_{\mu})\boldsymbol{\gamma}^{\mu}
\!+\!2P_{\mu}\boldsymbol{\gamma}^{\mu}\!-\!2m\mathbb{I}\right]\psi\!=\!0\label{D}
\end{eqnarray}
in which identity $\boldsymbol{\gamma}_{i}\boldsymbol{\gamma}_{j}\boldsymbol{\gamma}_{k}\!
=\!\boldsymbol{\gamma}_{i}\eta_{jk}\!-\!\boldsymbol{\gamma}_{j}\eta_{ik}\!+\!\boldsymbol{\gamma}_{k}\eta_{ij}\!+\!i\varepsilon_{ijkq}\boldsymbol{\pi}\boldsymbol{\gamma}^{q}$ was used and where $R_{\mu\nu}^{\phantom{\mu\nu}\nu}\!=\!R_{\mu}$ and $\frac{1}{2}\varepsilon_{\mu\alpha\nu\iota}R^{\alpha\nu\iota}\!=\!B_{\mu}$ were introduced. Multiplying on the left by $\overline{\psi}$ and selecting the imaginary part gives the real scalar condition
\begin{eqnarray}
(\nabla_{\mu}\ln{\phi^{2}}\!+\!R_{\mu})U^{\mu}\!=\!0
\end{eqnarray}
which can be worked out, with the help of (\ref{ds-du}), to give
\begin{eqnarray}
\nabla_{\mu}U^{\mu}\!=\!0;
\end{eqnarray}
if instead we select the real part, we obtain the condition of on-shell Lagrangian for the Dirac field
\begin{eqnarray}
&(\nabla_{\mu}\beta\!-\!2XW_{\mu}\!+\!B_{\mu})S^{\mu}\!+\!2P_{\mu}U^{\mu}\!-\!2m\Phi\!=\!0:
\end{eqnarray}
hence, we can say that if we contract on the left by $\overline{\psi}$ the imaginary and real parts give
\begin{eqnarray}
\overline{\psi}: \Big\{\begin{array}{cc}
\mathrm{Im}\ \longrightarrow & \nabla_{\mu}U^{\mu}\!=\!0\\
\mathrm{Re}\ \longrightarrow & (\nabla_{\mu}\beta\!-\!2XW_{\mu}\!+\!B_{\mu})S^{\mu}\!+\!2P_{\mu}U^{\mu}\!-\!2m\Phi\!=\!0.\\
\end{array}
\end{eqnarray}
If we multiply on the left by $\overline{\psi}\boldsymbol{\pi}$ the imaginary and real parts are
\begin{eqnarray}
\overline{\psi}\boldsymbol{\pi}: \Big\{\begin{array}{cc}
\mathrm{Im}\ \longrightarrow & \nabla_{\mu}S^{\mu}\!-\!2m\Theta\!=\!0\\
\mathrm{Re}\ \longrightarrow & (\nabla_{\mu}\beta\!-\!2XW_{\mu}\!+\!B_{\mu})U^{\mu}\!+\!2P_{\mu}S^{\mu}\!=\!0.\\
\end{array}
\end{eqnarray}
In an analogous manner, we can multiply by $\overline{\psi}\boldsymbol{\gamma}^{a}$, by $\overline{\psi}\boldsymbol{\gamma}^{a}\boldsymbol{\pi}$, or by $\overline{\psi}\boldsymbol{\sigma}^{ij}$, obtaining imaginary and real parts as
\begin{eqnarray}
\overline{\psi}\boldsymbol{\gamma}^{a}: \Big\{\begin{array}{cc}
\mathrm{Im}\ \longrightarrow & \nabla_{\alpha}\Phi
\!+\!(2XW_{\alpha}\!-\!B_{\alpha})\Theta\!+\!R_{\alpha}\Phi\!+\!2P^{\mu}M_{\mu\alpha}\!=0\\
\mathrm{Re}\ \longrightarrow & \nabla_{\mu}M^{\mu\alpha}\!-\!2XW_{\sigma}\Sigma^{\sigma\alpha}\!+\!\frac{1}{2}R^{ij\alpha}M_{ij}\!-\!2P^{\alpha}\Phi
\!+\!2mU^{\alpha}\!=\!0,\\
\end{array}\label{momentum}
\end{eqnarray}
\begin{eqnarray}
\overline{\psi}\boldsymbol{\gamma}^{a}\boldsymbol{\pi}: \Big\{\begin{array}{cc}
\mathrm{Im}\ \longrightarrow & \nabla^{\mu}\Sigma_{\mu\alpha}\!+\!2XW^{\mu}M_{\mu\alpha}
\!+\!\frac{1}{2}R_{ij\alpha}\Sigma^{ij}\!+\!2P_{\alpha}\Theta\!=\!0\\
\mathrm{Re}\ \longrightarrow & \nabla_{\nu}\Theta\!-\!(2XW_{\nu}\!-\!B_{\nu})\Phi
\!+\!R_{\nu}\Theta\!-\!2P^{\mu}\Sigma_{\mu\nu}\!+\!2mS_{\nu}\!=\!0,\\
\end{array}
\end{eqnarray}
\begin{eqnarray}
\overline{\psi}\boldsymbol{\sigma}^{ij}: \Big\{\begin{array}{cc}
\mathrm{Im}\ \longrightarrow & \nabla^{[\alpha}U^{\nu]}
\!+\!R^{[\alpha}U^{\nu]}\!-\!R^{\alpha\nu\mu}U_{\mu}
\!+\!\varepsilon^{\alpha\nu\mu\rho}(\nabla_{\mu}\beta\!-\!2XW_{\mu})U_{\rho}
\!+\!2\varepsilon^{\alpha\nu\mu\rho}P_{\mu}S_{\rho}\!-\!2mM^{\alpha\nu}\!=\!0\\
\mathrm{Re}\ \longrightarrow & \nabla_{\alpha}S_{\nu}\varepsilon^{\alpha\nu\sigma\mu}
\!+\!\frac{1}{2}R_{\eta\pi}^{\phantom{\eta\pi}\sigma}S_{\kappa}\varepsilon^{\eta\pi\mu\kappa}
\!-\!\frac{1}{2}R_{\eta\pi}^{\phantom{\eta\pi}\mu}S_{\kappa}\varepsilon^{\eta\pi\sigma\kappa}
\!-\!(\nabla\beta\!-\!2XW)^{[\sigma}S^{\mu]}
\!-\!2P^{[\sigma}U^{\mu]}\!=\!0.\\
\end{array}\label{curls}
\end{eqnarray}
These last ten equations are called Gordon decompositions \cite{Fabbri:2023onb}. As for the energy density, its polar form is given by
\begin{eqnarray}
\nonumber
&T^{\rho\sigma}\!=\!\frac{1}{4}F^{2}g^{\rho\sigma}
\!-\!F^{\rho\alpha}\!F^{\sigma}_{\phantom{\sigma}\alpha}+\\
\nonumber
&+\frac{1}{4}(\partial W)^{2}g^{\rho\sigma}
\!-\!(\partial W)^{\sigma\alpha}(\partial W)^{\rho}_{\phantom{\rho}\alpha}
\!+\!M^{2}(W^{\rho}W^{\sigma}\!-\!\frac{1}{2}W^{2}g^{\rho\sigma})+\\
&+\phi^{2}[P^{\rho}u^{\sigma}\!+\!P^{\sigma}u^{\rho}
\!+\!(\nabla^{\rho}\beta/2\!-\!XW^{\rho})s^{\sigma}
\!+\!(\nabla^{\sigma}\beta/2\!-\!XW^{\sigma})s^{\rho}
\!-\!\frac{1}{4}R_{\alpha\nu}^{\phantom{\alpha\nu}\sigma}s_{\kappa}
\varepsilon^{\rho\alpha\nu\kappa}
\!-\!\frac{1}{4}R_{\alpha\nu}^{\phantom{\alpha\nu}\rho}s_{\kappa}
\varepsilon^{\sigma\alpha\nu\kappa}].\label{energy}
\end{eqnarray}

Introducing the spin tensor as the Hodge dual of the spin axial-vector $2S^{\sigma\rho\alpha}\!=\!\varepsilon^{\sigma\rho\alpha\nu}S_{\nu}$, the second of (\ref{curls}) becomes
\begin{eqnarray}
&\nabla_{\alpha}S^{\alpha\sigma\mu}
\!+\!\frac{1}{2}R_{\eta\pi}^{\phantom{\eta\pi}[\sigma}S^{\mu]\eta\pi}
\!-\!\frac{1}{6}(\nabla\beta\!-\!2XW)^{[\sigma}\varepsilon^{\mu]\eta\rho\alpha}S_{\eta\rho\alpha}
\!-\!P^{[\sigma}U^{\mu]}\!=\!0\label{MPD}
\end{eqnarray}
which is just the conservation law of the spin tensor \cite{Fabbri:2023yhl}. The conservation law of the energy tensor comes directly from the Einstein gravitational field equations and it reads $\nabla_{\alpha}T^{\alpha\nu}\!\equiv\!0$, so that when (\ref{energy}) is plugged, we obtain
\begin{eqnarray}
&U^{\sigma}\nabla_{\sigma}P^{\rho}\!=\!qF^{\rho\alpha}U_{\alpha}
\!+\!X\nabla^{\rho}W^{\alpha}S_{\alpha}
\!+\!\frac{1}{2}\nabla^{\sigma}(-S_{\sigma}\nabla^{\rho}\beta
\!+\!S_{\sigma\alpha\nu}R^{\alpha\nu\rho})\label{Newton}
\end{eqnarray}
in which the torsion and electrodynamic field equations have also been used.

This last equation (\ref{Newton}) is the Newton equation of motion for matter distributions, or more specifically one may also refer to it as Navier-Stokes equation of motion for fluids. It can also be seen as the linear Mathisson-Papapetrou-Dixon equation in parallel to (\ref{MPD}) being the angular Mathisson-Papapetrou-Dixon equation.

In the present paper, we shall refer to the system of equations (\ref{MPD}-\ref{Newton}) as the Mathisson-Papapetrou-Dixon equations.

Here, they have been obtained from the Dirac spinor field equation, in the most general background accounting for electrodynamics, curvature and torsion, with torsion taken as a propagating Proca field.
\section{Two Limiting Conditions}\label{special}
\subsection{Effective Approximation}
The above system of Mathisson-Papapetrou-Dixon equations (\ref{MPD}-\ref{Newton}) is for now fully general.

Nevertheless, there are special conditions that are worth studying. One is about torsion, and in particular the fact that in this model it is a propagating massive axial-vector field. In the situation in which the mass of torsion were to be very large, it could make sense to consider the effective approximation, that is the regime in which all derivative terms are negligible compared to massive terms. When this condition is implemented, we have that the torsion field equations (\ref{torsion}) reduce to
\begin{eqnarray}
&M^{2}W^{\mu}\!=\!XS^{\mu}
\end{eqnarray}
telling that torsion is proportional to the spin axial-vector. As such, torsion can be substituted everywhere in terms of the spin axial-vector itself. The resulting effective theory will be one in which non-linear terms will appear.

The MPD equations (\ref{MPD}-\ref{Newton}), with the above hypothesis, become
\begin{eqnarray}
&\nabla_{\alpha}S^{\alpha\sigma\mu}
\!+\!\frac{1}{2}R_{\eta\pi}^{\phantom{\eta\pi}[\sigma}S^{\mu]\eta\pi}
\!-\!\frac{1}{6}\nabla^{[\sigma}\beta\varepsilon^{\mu]\eta\rho\alpha}S_{\eta\rho\alpha}
\!-\!P^{[\sigma}U^{\mu]}\!=\!0\label{MPDEff}\\
&U^{\sigma}\nabla_{\sigma}P^{\rho}\!=\!qF^{\rho\alpha}U_{\alpha}
\!+\!\nabla^{\sigma}(p\delta^{\rho}_{\sigma}
\!-\!\frac{1}{2}S_{\sigma}\nabla^{\rho}\beta
\!+\!\frac{1}{2}S_{\sigma\alpha\nu}R^{\alpha\nu\rho})\label{NewtonEff}
\end{eqnarray}
where all instances of torsion have disappeared from the angular MPD equation (\ref{MPDEff}) and they have been reduced to a pure pressure term $p\!=\!-2\phi^{4}X^{2}/M^{2}$ in the linear MPD equation (\ref{NewtonEff}). This pressure term is non-linear.

Notice that with respect to the standard MPD equations there are also two deviations coming from the interaction of the spin tensor $S_{\eta\rho\alpha}$ with the gradient of the chiral angle $\nabla_{\sigma}\beta$ and with the torsionless space-time tensorial connection $R_{\eta\pi\sigma}$: hence, the usual MPD equations can only be obtained if $\nabla_{\sigma}\beta\!=\!R_{\eta\pi\sigma}\!=\!0$ hold.
\subsection{Plane Waves}
We want to study now the condition of plane waves. In any quantum field theoretical treatments, this condition is implemented by the requirement that the spinor be such that
\begin{eqnarray}
&i\boldsymbol{\nabla}_{\mu}\psi\!=\!P_{\mu}\psi\label{pw}.
\end{eqnarray}
This is also the assumption made in \cite{Guedes:2022mdy} to obtain the Mathisson-Papapetrou-Dixon equations.

Writing the above as $\boldsymbol{\nabla}_{\mu}\psi\!=\!-iP_{\mu}\psi$, we can compare it against the general expression of covariant derivative (\ref{decspinder}): we see that one must have
\begin{eqnarray}
&\nabla_{\mu}\ln{\phi}\mathbb{I}
\!-\!\frac{i}{2}\nabla_{\mu}\beta\boldsymbol{\pi}
\!-\!\frac{1}{2}R_{ij\mu}\boldsymbol{\sigma}^{ij}\!=\!0
\end{eqnarray}
and therefore
\begin{eqnarray}
&\nabla_{\mu}\phi\!=\!\nabla_{\mu}\beta\!=\!R_{ij\mu}\!=\!0
\end{eqnarray}
due to the linear independence of the matrices. As $\nabla_{\mu}\beta\!=\!0$ means that $\beta$ is constant, and since every pseudo-scalar constant must necessarily vanish, we have that $\beta\!=\!0$ must hold. Nevertheless, this condition is implied by the other two. In fact, enforcing $R_{ab\mu}\!=\!\nabla_{\mu}\phi\!=\!0$ onto (\ref{ds-du}) and using the definition of the spin axial-vector, we obtain $\nabla_{\mu}S^{\mu}\!=\!0$, so the Gordon decomposition $\nabla_{i}S^{i}\!=\!2m\Theta$ gives $\Theta\!=\!0$ identically: this means that $\beta\!=\!0$ is implied by $R_{ab\mu}\!=\!\nabla_{\mu}\phi\!=\!0$ in general. Consequently, the conditions
\begin{eqnarray}
&\nabla_{\mu}\phi\!=\!R_{ij\mu}\!=\!0\label{pwpolar}
\end{eqnarray}
account for the smallest set of conditions equivalent to the plane wave structure of the spinor.

The MPD equations (\ref{MPDEff}-\ref{NewtonEff}), when the plane wave conditions are implemented, reduce to
\begin{eqnarray}
&\nabla_{\alpha}S^{\alpha\sigma\mu}\!=\!P^{[\sigma}U^{\mu]}\label{MPDEffpw}\\
&u^{\sigma}\nabla_{\sigma}P^{\rho}\!=\!qF^{\rho\alpha}u_{\alpha}\label{NewtonEffpw}
\end{eqnarray}
showing that (\ref{MPDEffpw}-\ref{NewtonEffpw}) have indeed the form of the MPD equations found in \cite{Guedes:2022mdy}.

To be precise, the above are the MPD equations in the form found in \cite{Guedes:2022mdy} after that all the torsionfull quantities are decomposed in torsionless quantities plus torsional contributions by employing the identities (\ref{decV}-\ref{decSpin}, \ref{decG}).
\section{Trivial Reduction}
In section \ref{general} we have obtained the MPD equations in general, and in section \ref{special} we have recovered them with the special form in which they have been obtained in \cite{Guedes:2022mdy}. However, let us see where the plane wave condition will take us.

When the plane wave conditions (\ref{pwpolar}) are implemented, it is an easy exercise to prove that all spinor bi-linears turn out to be covariantly constant. In fact, because $\beta\!=\!0$ and $\phi$ is constant then $\Theta\!=\!0$ and $\Phi$ is constant, and as $R_{ab\mu}\!=\!0$ then (\ref{ds-du}) imply that $s_{i}$ and $u_{i}$ are also covariantly constant.

Of the Gordon decompositions (\ref{momentum}), the second reduces to
\begin{eqnarray}
&P^{\alpha}\Phi\!=\!mU^{\alpha}\!-\!XW_{\sigma}\Sigma^{\sigma\alpha};
\end{eqnarray}
in effective approximation, we have that $XW_{\sigma}\Sigma^{\sigma\alpha}\!=\!X^{2}/M^{2}S_{\sigma}\Sigma^{\sigma\alpha}\!\equiv\!X^{2}/M^{2}\Phi U^{\alpha}$ because of (\ref{products-2}): hence
\begin{eqnarray}
&P^{\alpha}\!=\!(m\!-\!2\phi^{2}X^{2}/M^{2})u^{\alpha}.
\end{eqnarray}
This means that momentum and velocity are proportional. As $\phi$ and $u_{i}$ are covariantly constant, it also means that the momentum is covariantly constant. And because of (\ref{Maxwell}), we have $F_{\mu\nu}\!=\!0$ identically. Finally we have that $\phi$ and $s_{i}$ are covariantly constant, so $S_{i}$ is covariantly constant and thus $S_{abk}$ is covariantly constant. As a consequence, we have that the MPD equations (\ref{MPDEffpw}-\ref{NewtonEffpw}) reduce to the form
\begin{eqnarray}
&0\!=\!0\\
&0\!=\!0
\end{eqnarray}
and as it is now evident these equations become void of any physical content.
\section{Conclusion}
In this paper, we have considered spinor fields in polar form, developing their dynamical features on manifolds with curvature and torsion. The dynamics has been established by a Lagrangian in which the spinor interacted with electrodynamics, beside gravity and torsion, and where torsion was taken in its most general form compatible with the restrictions of being a Proca axial-vector field. All field equations were derived, and from the spinor field equations we also derived the ten Gordon decompositions. The latter of these Gordon decompositions, that is the conservation law for the spin, was recognized as the angular Mathisson-Papapetrou-Dixon equation, while the conservation law of the energy was the Newton equation of hydrodynamic motion, that is the linear Mathisson-Papapetrou-Dixon equation: the angular and linear MPD equations have thus been obtained in the most general case of propagating torsion gravity.

We have then considered two common approximations, that is torsion in effective approximation and the spinor in plane-wave form. We have shown however that with these two restrictions, both MPD equations trivialize.

Because from the Dirac theory (which is the essence of relativistic quantum mechanics) it is possible to derive the Mathisson-Papapetrou-Dixon equations (which are the basic equations of motion of spin fluids), then it is possible in quantum mechanics to have at least some quantities described at least in part by classical characters.

\

\textbf{Funding and acknowledgements}. This work was funded by Next Generation EU project ``Geometrical and Topological effects on Quantum Matter (GeTOnQuaM)''.

\

\textbf{Conflict of interest}. The author declares no conflict of interest.

\end{document}